\documentclass[11pt,a4paper]{article}
\pdfoutput=1

\usepackage{jcappub}

\usepackage{amsmath}
\usepackage{amssymb}
\usepackage{epsfig}
\usepackage{graphicx}
\usepackage{multirow}
\usepackage{longtable}
\usepackage{lscape}
\usepackage{bbm}
\usepackage{dcolumn}
\usepackage{rotating}
\usepackage[squaren]{SIunits}
\usepackage[T1]{fontenc} % if needed

\allowdisplaybreaks[4] % for page break

\newcommand{\capdef}{}
\newcommand{\mycaption}[2][\capdef]{\renewcommand{\capdef}{#2}%
        \caption[#1]{{\footnotesize #2}}}
\makeatletter
\renewcommand{\fnum@table}{\textbf{\tablename~\thetable}}
\renewcommand{\fnum@figure}{\textbf{\figurename~\thefigure}}
\makeatother

\newcounter{myenumi}

\renewcommand{\themyenumi}{\roman{myenumi}}
{\end{list}}

\newlength{\myem}
\newcounter{mysubequation}[equation]

\makeatletter
\renewcommand{\section}{\@startsection{section}{1}{0em}{-\baselineskip}%
{\baselineskip}{\normalfont\large\bfseries}}
\renewcommand{\subsection}%
{\@startsection{subsection}{2}{0em}{-0.7\baselineskip}%
{0.7\baselineskip}{\normalfont\bfseries}}
\makeatother

\newcommand{\bi}{\begin{itemize}}
\newcommand{\ei}{\end{itemize}}

\newcommand{\be}{\begin{equation}}
\newcommand{\ee}{\end{equation}}
\newcommand{\bea}{\begin{eqnarray}}
\newcommand{\eea}{\end{eqnarray}}

\newcommand{\deltacp}{\delta_{\mathrm{CP}}}

\newcommand{\ie}{{\it i.e.}}

\newcommand{\eg}{{\it e.g.}}

\newcommand{\cf}{{\it cf.}}

\newcommand{\eq}{Eq.}

\newcommand{\fig}{Fig.}

\newcommand{\Ref}{Ref.}
\newcommand{\Refs}{Refs.}
\newcommand{\Sec}{Sec.}

\newcommand{\Tab}{Table}

\newcommand{\equ}[1]{\eq~(\ref{equ:#1})}
\newcommand{\figu}[1]{\fig~\ref{fig:#1}}

\title{\textbf{Neutrino decays over cosmological distances and the implications for neutrino telescopes}}

\author{Philipp~Baerwald,}
\author{Mauricio~Bustamante,}
\author{and Walter~Winter}
\affiliation{Institut f{\"u}r Theoretische Physik und Astrophysik, \\ Universit{\"a}t W{\"u}rzburg, 
       97074 W{\"u}rzburg, Germany}

\emailAdd{philipp.baerwald@physik.uni-wuerzburg.de}
\emailAdd{mauricio.bustamante@physik.uni-wuerzburg.de}
\emailAdd{winter@physik.uni-wuerzburg.de}

\abstract{We discuss decays of ultra-relativistic neutrinos over cosmological distances by solving the decay equation in terms of its redshift dependence. We demonstrate that there are significant conceptual differences compared to more simplified  treatments of neutrino decay. For instance, the maximum distance the neutrinos have traveled is limited by the Hubble length, which means that the common belief that longer neutrino lifetimes can be probed by longer distances does not apply. As a consequence, the neutrino lifetime limit from supernova 1987A cannot be exceeded by high-energy astrophysical neutrinos. We discuss the implications for neutrino spectra and flavor ratios from gamma-ray bursts as one example of extragalactic sources, using up-to-date neutrino flux predictions. 
If the observation of SN 1987A implies that $\nu_1$ is stable and the other mass eigenstates decay with rates much smaller than their current bounds, the muon track rate can be substantially suppressed compared to the cascade rate in the region IceCube is most sensitive to. In this scenario, no gamma-ray burst neutrinos may be found using muon tracks even with the full scale experiment, whereas reliable information on high-energy astrophysical sources can only be obtained from cascade measurements. As another consequence, the recently observed two cascade event candidates at PeV energies will not be accompanied by corresponding muon tracks.}


\arxivnumber{1208.4600}

\begin{document}
\maketitle
\flushbottom

\section{Introduction}

While the masses and mixings of the neutrinos are already well probed, other  neutrino properties are less understood, such as the  electromagnetic properties and the neutrino lifetime.   The most stringent phenomenological bound on neutrino lifetime comes from the observation of neutrinos from supernova 1987A~\cite{Hirata:1987hu,Bionta:1987qt}, which were measured in the electron flavor. Given the uncertainty on the supernova neutrino flux (order 50\%) and the neutrino mixing parameters, it may apply to the mass eigenstate $\nu_1$ or $\nu_2$. For the sake of simplicity, we assume that the bound applies to $\nu_1$: $\tau_1/m_1 \gtrsim
10^5~\text{s/eV}$.\footnote{Neutrino lifetime is usually described by  $\kappa_i^{-1} \equiv \tau_{i,0}/m_i$, where $\tau_{i,0}$ is the rest frame lifetime of the  mass eigenstate $\nu_i$. The origin of the mass dependence is the fact that decays scale as $\exp[-t /
(\tau_{i,0} \gamma)] \simeq \exp[-(L m_i) / (E \tau_{i,0})]$, \ie, the rest frame lifetime $\tau_{i,0}$ is  boosted by $\gamma
= E/m_i$ into the observer's frame, and $L$ (baseline source-detector) and $E$ (neutrino energy) are quantities related to source and experiment.} The (model-independent) bounds on the other mass eigenstates are less stringent:
bounds on $\nu_2$ lifetime are imposed
by solar neutrino data, yielding $\tau_2/m_2 \gtrsim
10^{-4}~\text{s/eV}$ for decays into invisible daughters~\cite{Joshipura:2002fb,Bandyopadhyay:2002qg,Beacom:2002cb} and $\tau_2/m_2 \gtrsim 10^{-3}~\text{s/eV}$ for
decay modes with secondary $\bar{\nu}_e$
appearance~\cite{Eguchi:2003gg, Aharmim:2004uf}. Furthermore, $\nu_3$ is constrained from the analysis of atmospheric and long-baseline
neutrino data, $\tau_3/m_3 \gtrsim
10^{-10}~\text{s/eV}$~\cite{GonzalezGarcia:2008ru}. More stringent bounds can be derived  when specific decay models are assumed (see, \eg,
\Refs~\cite{Pakvasa:2003db, Yao:2006px, Pakvasa:2008nx} for an
overview). For example, solar neutrinos strongly limit the possibility
of radiative decays~\cite{Raffelt:1985rj}, while for Majoron
decays~\cite{Gelmini:1980re, Chikashige:1980qk}, explicit bounds can be
obtained from neutrino-less double-beta decay and
supernovae~\cite{Tomas:2001dh}. Another possibility is the decay into un-particles~\cite{Chen:2007zy, Zhou:2007zq, Li:2007kj, Majumdar:2007mp}. In this study, we do not consider specific decay models, but focus on the phenomenology of neutrino decay, given the bounds on the lifetimes above. Especially neutrino telescopes~\cite{Aslanides:1999vq, Ahrens:2002dv, Tzamarias:2003wd, Piattelli:2005hz} are sensitive to neutrinos with an average energy and traveled distance many orders of magnitude larger
than present neutrino experiments, and may be an interesting approach to probe neutrino decay.

Neutrino oscillations are typically assumed to be averaged out over cosmological distances (see \Ref~\cite{Farzan:2008eg} for a discussion), and the usual flavor mixing remains to be taken into account. 
As far as decays are concerned, one distinguishes between decays into products {\em invisible} to the detector, such as sterile neutrinos, un-particle states, Majorons, or active neutrinos strongly degraded in energy, and decays into {\em visible} states, \ie, active neutrino flavors. The decays can be {\em complete}, \ie, all unstable mass eigenstates have decayed (see, \eg, \Ref~\cite{Beacom:2002vi}), or {\em incomplete}, \ie, the decay spectral signature will be visible. A complete classification of complete (visible or invisible) decay scenarios has been performed in \Ref~\cite{Maltoni:2008jr}, while incomplete invisible decay has, for instance, been studied for active galactic nuclei (AGNs) in \Ref~\cite{Bhattacharya:2009tx,Bhattacharya:2010xj}. The description of incomplete visible decays is, in general, more complicated~\cite{Lindner:2001fx, Lindner:2001th}, which is why we focus on incomplete invisible decay in this study (IC-40 refers to the 40-string configuration of the IceCube detector). The characteristic energy dependence of neutrino decay may lead to easily observable imprints in the neutrino fluxes in that case~\cite{Bhattacharya:2009tx,Bhattacharya:2010xj}. On the other hand, it has been proposed to use the flavor composition at the detector, such as the ratio between muon tracks and cascades, to distinguish different decay scenarios~\cite{Beacom:2002vi, Lipari:2007su, Majumdar:2007mp, Maltoni:2008jr,Bustamante:2010nq,Mehta:2011qb}.
We will discuss both options in this study, where our main focus are the fluxes.

It is often believed that decays are always complete if the sources lie far enough away, which, in turn, allows for the test of very long neutrino lifetimes of the order of
\begin{equation}
 \kappa^{-1} \, \left[ \frac{\vphantom{M} \mathrm{s}}{\mathrm{eV}} \right] \equiv 
\frac{\tau \, [\mathrm{s}]}{m \, [\mathrm{eV}]}
\simeq 10^2 \, \frac{L \, [\mathrm{Mpc}]}{E \, [\mathrm{TeV}]} \, . \label{equ:est}
\end{equation}
As a consequence, especially objects with high redshifts may be well suited to test neutrino lifetime, and may potentially lead to bounds stronger than the one from SN 1987A. An example, which we are going to use in this work, are gamma-ray bursts (GRBs), with observed redshifts as high as $z \simeq 6-8$; see \Ref~\cite{Kistler:2009mv} for the redshift and luminosity distribution. This leads to potentially strong constraints via \equ{est}. However, note that the relationship between distance $L$ and redshift $z$ depends for $z \gtrsim 0.1$ (or $L \gtrsim 360 \, \mathrm{Mpc}$) on the cosmological distance measure to be used, which we are going to address in this study; see \Refs~\cite{Wagner:2011uy,Weiler:1994hw} for a discussion in the context of neutrino oscillations, and \Refs~\cite{Beacom:2003eu,Esmaili:2012he} in the context of pseudo-Dirac neutrinos.
For the neutrino flux, especially redshifts $z \simeq 1$ will dominate~\cite{Baerwald:2011ee}, which is a consequence of the convolution of the star formation rate (including some redshift evolution function) and the contribution to the total flux scaling as $1/d_L^2$, where $d_L$ is the luminosity distance. This value is already significantly beyond the indicated $z \simeq 0.1$, which means that cosmological effects have to be taken into account. 

GRBs are not only an interesting class of candidate neutrino sources because they may originate from high redshifts, but also because the non-observation of neutrinos from GRBs~\cite{Abbasi:2012zw} has started to constrain the gamma-ray--neutrino connection in conventional internal shock models~\cite{Guetta:2003wi,Abbasi:2009ig}, based on the ideas in \Ref~\cite{Waxman:1997ti}. A re-calculation of the predicted neutrino fluence from the gamma-ray fluence has yielded an order of magnitude lower result~\cite{Hummer:2011ms} for the same burst parameters (see also \Ref~\cite{Li:2011ah}), which is a possible explanation soon going to be tested by new data. We will use this new nominal prediction for the quasi-diffuse flux~\cite{Hummer:2011ms}, calculated with the IC-40 bursts~\cite{Abbasi:2011qc}, in this study. Note that model-specific~\cite{Hummer:2011ms,Barranco:2012xj} and more generic~\cite{He:2012tq} astrophysical uncertainties may imply even lower neutrino flux predictions. On the other hand, another plausible explanation for why nothing has been seen yet may be that neutrinos have actually (at least partially) decayed between source and detector. We will investigate the impact of this hypothesis on the predicted fluxes. 

One of the very recent  puzzles of neutrino astronomy is the potential observation of two cascade events at IceCube at  PeV energies~\cite{NeutrinoTalk}. In particular, assuming that these are of extragalactic origin, an interesting question is: Why have these been observed as cascades, and no muon tracks have been seen? This question may become more prominent with increasing statistics. In this work, we demonstrate that neutrino decay can provide an answer to that question, given the current constraints on neutrino lifetime; for alternative explanations, see \Refs~\cite{Barger:2012mz, Fargion:2012zc}. We will demonstrate that muon tracks may be strongly suppressed in the presence of neutrino decay compared to (especially electromagnetic) cascades, and that cascade measurements are much more powerful to find astrophysical neutrinos in this case. We will use GRBs as an example, but our conclusions can be applied to AGNs as well.

This study is organized as follows: We describe the decay framework in 
\Sec~\ref{sec:decay}, which consists of the general framework including cosmological distances, a simplified solution, and the proper solution of the redshift-dependent decay equation. We then discuss the implications for GRBs in \Sec~\ref{sec:grb}, as one possible example of extragalactic neutrino sources. In that section, we introduce several benchmark GRBs and the analysis techniques, discuss the impact of decay on the neutrino fluxes, and introduce a phenomenologically viable model for which we show the implications for fluxes and flavor composition. In \Sec~\ref{sec:icecube} we apply this model to a realistic stacking analysis of IceCube, and compare the predicted neutrino fluxes to current and future bounds. Finally, we conclude in 
\Sec~\ref{sec:summary}. Note that, while we use GRBs as test case, the results from this study apply to the neutrino propagation from other extragalactic sources as well, including cosmogenic neutrinos.

\section{Decay framework}
\label{sec:decay}

Here we introduce in the first subsection the general decay framework including the aspect of cosmological distances. We solve the decay equation in a simplified approach in the second subsection to study when a cosmological framework has to be used, and we show the proper solution in the third subsection.

\subsection{General framework including cosmology}

Neutrino decay of the mass eigenstate $\nu_i$ can be described by the usual
differential equation
\begin{equation}
 \label{equ:EqDecay}
 \frac{dN_i}{dt} = - \lambda_i N_i  \, .
\end{equation}
 Here the decay rate (inverse lifetime) $\lambda_i$ is given by
\begin{equation}\label{equ:EqDefLambdaSimple}
 \lambda_i = \frac{m_i}{\tau_{i,0}} \frac{1}{E} \equiv \frac{\kappa_i}{E} \, ,
\end{equation}
with $\tau_{i,0}$ the rest frame lifetime boosted by $\gamma_i=E/m_i$ into the observer's  frame, $\kappa_i \equiv m_i/\tau_{i,0}$, and $E$ the neutrino energy. Since neither the rest frame lifetime $\tau_{i,0}$ nor the lifetime $\tau_i=\gamma_i \tau_{i,0}$  are directly observable, typically $\kappa_i^{-1} =\tau_{i,0}/m_i$ $[\mathrm{s \, eV^{-1}}]$ is quoted as the lifetime of the neutrino mass eigenstate $\nu_i$.

The time $t$ is for ultra-relativistic neutrinos typically associated with the baseline $L$, which is the distance between source and detector. Let us assume that the neutrinos under study are produced at a source with a redshift $z$. Over cosmological distances, we therefore need to express this distance in terms of redshift, \ie, $L(z)$. In cosmology, however, the relationship between distance and redshift depends on the cosmological distance measure, which can be luminosity distance, angular diameter distance, proper distance, or light-travel distance; see, \eg, \Ref~\cite{Hogg:1999ad}. For neutrino decay, the relevant quantity is the time traveled in \equ{EqDecay}, which is to be associated with the light-travel or look-back distance. In other words, the clock which triggers neutrino decay is directly related to the distance the neutrino has traveled since its production.  By definition,
\begin{equation}\label{equ:EqLDef}
 L\left(z\right) = L_H \int_0^z \frac{dz^\prime}{\left(1+z^\prime\right) h\left(z^\prime\right)} ~,
\end{equation}
with $L_H \equiv c/H_0 \approx 3.89$~Gpc the Hubble length \cite{Nakamura:2010zzi}, $h\left(z\right) \equiv H\left(z\right)/H_0$, and $H_0$ the Hubble constant. The Hubble parameter $H(z)$ is defined as
\begin{equation}
 H\left(z\right) \equiv H_0 \sqrt{\Omega_m \left(1+z\right)^3 + \Omega_\Lambda} \, ,
\end{equation}
assuming a flat $\Lambda$CDM cosmology. We choose $\Omega_m = 0.27$ and $\Omega_\Lambda = 0.73$ in the following \cite{Komatsu:2010fb}, unless explicitly noted otherwise.

\begin{figure}[t!]
 \centering
 \includegraphics[width=0.5\textwidth]{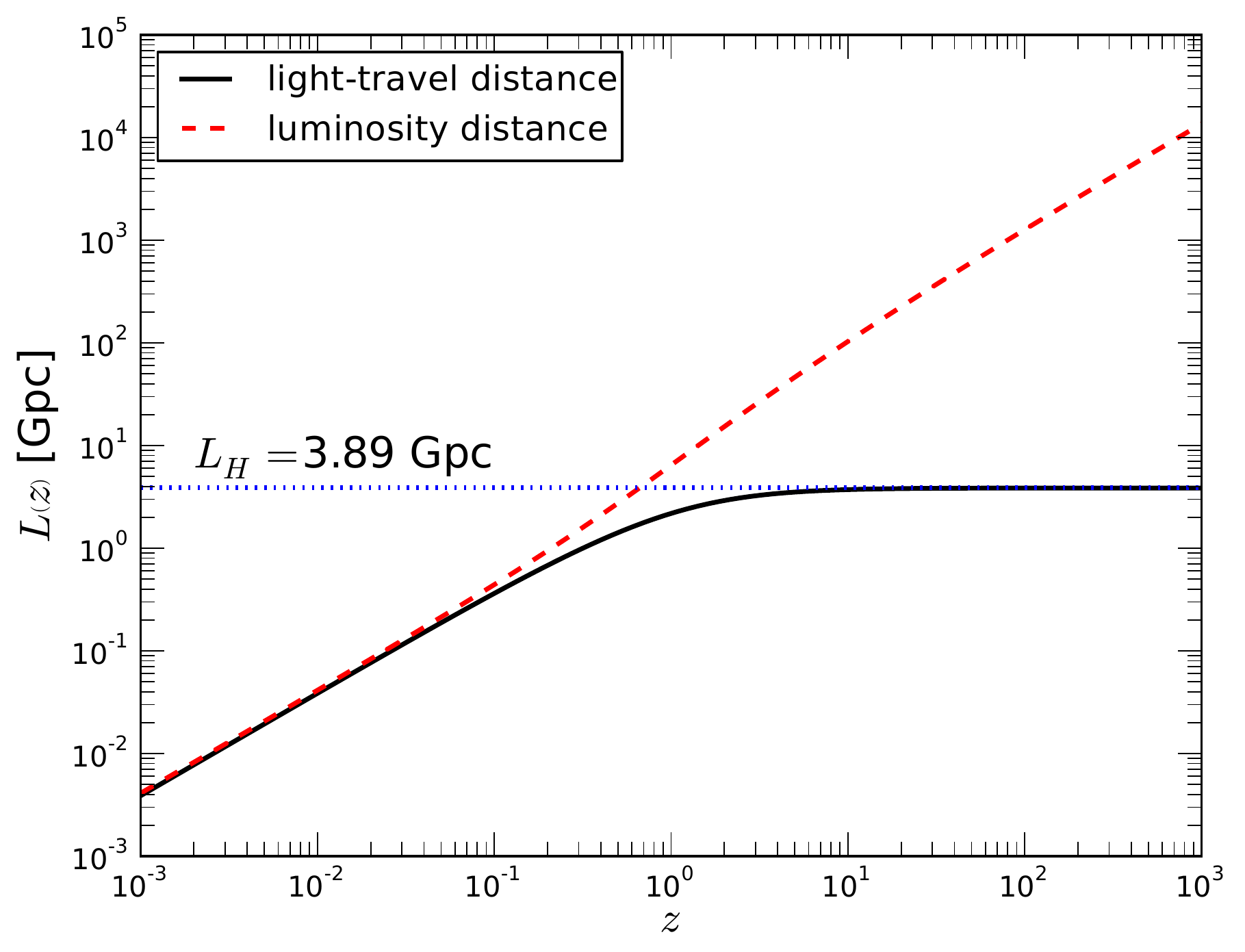}
 \mycaption{\label{fig:FigLvsZ}Light-travel distance as a function of redshift in a $\Lambda$CDM cosmology (solid black line). For comparison, we have also included the luminosity distance (dashed curve).}
\end{figure}

The light-travel distance $L(z)$ is shown in \figu{FigLvsZ} compared to the luminosity distance. It is limited by the Hubble length $L_H \simeq 3.89$~Gpc, which is the horizon beyond which ultra-relativistic particles cannot be seen. This has interesting implications for neutrino decay: the maximum distance relevant for the test of decay is limited, and therefore also the maximum time the ultra-relativistic neutrinos have traveled. From \figu{FigLvsZ} it is clear that the notion that larger distances can probe longer lifetimes only holds for $z \lesssim 0.1$ (or $L \lesssim 360 \, \mathrm{Mpc}$), where the cosmological distance measures are very similar. On the other hand, for $z \gtrsim 1$, the light-travel distance is limited. For ``typical'' neutrino peak energies higher than about $10^2 \, \mathrm{TeV}$ for GRBs and AGNs, one can estimate that the maximal testable lifetimes are about $\kappa^{-1} \sim 10^4 \, \mathrm{s \, eV^{-1}}$ from \equ{est}. Therefore, the lifetime limit from SN 1987A cannot be exceeded for the electron flavor (or the corresponding mass eigenstates). Although neutrino telescopes may also probe $\kappa_3^{-1}$  using muon tracks (the poorly constrained lifetime of $\nu_3$), any future galactic supernova neutrino observation, which may use different flavors and neutral currents, will likely provide much better bounds because of the lower neutrino energies.

The observable neutrino flux $\phi_\alpha$ ($\mathrm{GeV^{-1} \, cm^{-2} \, s^{-1}}$) including flavor mixing and decay (neglecting possible Earth attenuation effects) is given by
\begin{equation}\label{equ:EqDefFlavCompEarth}
 \phi_{\nu_\alpha+\bar{\nu}_\alpha}\left(E_0,z\right) = \sum_{\beta=e,\mu,\tau} P_{\alpha \beta}\left(E_0,z\right) \phi^0_{\nu_\beta+\bar{\nu}_\beta} \ , 
\end{equation}
where $\phi^0_{\nu_\beta+\bar{\nu}_\beta}$ is the neutrino flux without propagation effects and $E_0$ is the observed energy. Note that we simply add the neutrino and antineutrino fluxes here, since most observables are not sensitive to the neutrino polarity. The transition probability  $P_{\alpha \beta}$ for the transition $\nu_\alpha \rightarrow \nu_\beta$ including flavor mixing and decay is given by
\begin{equation}
 P_{\alpha\beta}(E_0,z) = \sum_i \left\vert U_{\alpha i} \right\vert^2 \left\vert U_{\beta i} \right\vert^2   \, \frac{N_i(E_0,z)}{\hat N_i(E_0)} =  \sum_i \left\vert U_{\alpha i} \right\vert^2 \left\vert U_{\beta i} \right\vert^2   \, D_i(E_0,z) \, ,
\label{equ:p}
\end{equation}
where $N_i(E_0,z)$ is the number of neutrinos left after traveling over a distance $L(z)$, $\hat N(E_0)$ is the initial number of neutrinos (as a function of the observed energy), and $D_i(E_0,z) \equiv N_i(E_0,z)/\hat N_i(E_0) \le 1$ is the corresponding damping factor~\cite{Blennow:2005yk}. It is the solution of \equ{EqDecay}, which we will address in the next subsections. 
 For the mixing angles, we use the best-fit values (for normal hierachy) from Ref.~\cite{Fogli:2012ua}, \ie, $\sin^2 \theta_{12} = 0.307$, $\sin^2 \theta_{23} = 0.398$, $\sin^2 \theta_{13} = 0.0245$, and the CP-violating phase $\delta = 0.89 \cdot \pi$.
As observables, we use the muon neutrino flux $\phi_\mu \equiv  \phi_{\nu_\mu+\bar{\nu}_\mu}$, representing muon tracks, the electron neutrino flux $\phi_e \equiv  \phi_{\nu_e+\bar{\nu}_e}$, representing electromagnetic cascades, and the flavor ratio $R \equiv \phi_\mu/(\phi_e + \phi_\tau)$~\cite{Serpico:2005sz}, representing the ratio between muon tracks and cascades.\footnote{For the sake of simplicity, we ignore neutral current cascades here, which would enter as a background.} The flavor composition at the source is computed in a self-consistent way including the cooling of the secondary pions, muons, and kaons, and the helicity-dependent muon decays; see, \eg, \Refs~\cite{Lipari:2007su,Kachelriess:2007tr,Hummer:2010ai,Winter:2012xq}. For example, for GRBs, a transition from $\phi^0_e:\phi^0_\mu:\phi^0_\tau = 1:2:0$ (pion decay source) to $0:1:0$ (muon-damped source) is expected, which can be predicted as a function of the source parameters.
 
\subsection{A simplified solution of the decay equation, and a stability paradox}
\label{sec:simple}

Here we solve the decay equation in the usual way used in the literature, and apply it to cosmological distances. Although we will demonstrate that this approach does not work, it is quite illustrative to understand the problems and in what redshift range simplified approaches can actually be used.

The decay equation, \equ{EqDecay}, has, for $\lambda_i = const.$, the well-known solution
\begin{equation}
N_i(t)=N_i(t=0) \, e^{-\lambda_i t} \, = \hat N_{i} \, e^{-\lambda_i t} \, , \label{equ:Nsimple}
\end{equation}
where $\hat N_{i}$ is the initial number of $\nu_i$.
We can easily modify that equation, using \equ{EqLDef}, such that\footnote{We use natural units, \ie, $\hbar=c=1$.}
\begin{equation}
N_i(z)=\hat N_{i} \, e^{-\lambda_i L(z)} \, , \label{equ:Nsimple2}
\end{equation}
where $N_i(z)$ is the number of remaining neutrinos after traveling $L(z)$. Note that here the redshift is measured relative to the origin $z$, 
whereas in cosmology $z=0$ (and quantities marked with ``0'') refer to the current epoch.

We also need to take into account the effect of the adiabatic expansion of the Universe on the neutrino energy: if $E(z)$ is the energy in the production epoch and $E_0$ is the observed energy, then
\begin{equation}\label{equ:EqEnergyRel}
 E(z) = \left(1+z\right) E_0 \, .
\end{equation}
With this, the decay rate in \equ{EqDefLambdaSimple} acquires a redshift dependence, \ie,
\begin{equation}\label{equ:EqDefLambda}
 \lambda_i = \lambda_i\left(z\right) = \frac{\kappa_i}{E_0\left(1+z\right)} \, .
\end{equation}
Using that in \equ{Nsimple2}, 
we can therefore write the damping factor  needed for \equ{p} as
\begin{equation}
D_i(E_0,z) =  \exp\left( -\frac{\kappa_i}{E_0} \frac{L\left(z\right)}{\left(1+z\right)} \right)  
\label{equ:decaysimp}
\end{equation}
with the definition of $L(z)$ from \equ{EqLDef}. In order to isolate the redshift-dependent part of the evolution, we define the dimensionless function
\begin{equation}
 \mathcal{Z}_1\left(z\right) \equiv \exp\left(\frac{L(z)}{L_H \cdot (1+z)}\right) \, ,
\label{equ:z1}
\end{equation}
such that \equ{decaysimp} can be re-written as
\begin{equation}
 \label{equ:EqNz1}
 D_i(E_0,z)  =  \left[\mathcal{Z}_1\left(z\right)\right]^{-\frac{\kappa_i L_H}{E_0}}  \, .
\end{equation}
Here the exponent depends on  $\kappa_i$ and observed energy $E_0$ only, whereas the base depends on redshift only. One can easily see that for $E_0 \rightarrow \infty$, the damping factor $D_i(E_0,z)  \rightarrow 1$, which corresponds to stable neutrinos. For finite $E_0$ and large decay rates $\kappa_i \gg E_0/L_H$, however, one finds $D_i(E_0,z) = 1$ (stable neutrinos) for $\mathcal{Z}_1=1$, $D_i(E_0,z) \rightarrow 0$ (complete decays) for $\mathcal{Z}_1>1$, and  $D_i(E_0,z) \rightarrow \infty$ (exploding solution) for $\mathcal{Z}_1<1$. Because large enough decay rates  correspond to complete decays, only the second case can be a physical solution.

\begin{figure}[tp]
 \centering
 \includegraphics[width=0.5\textwidth]{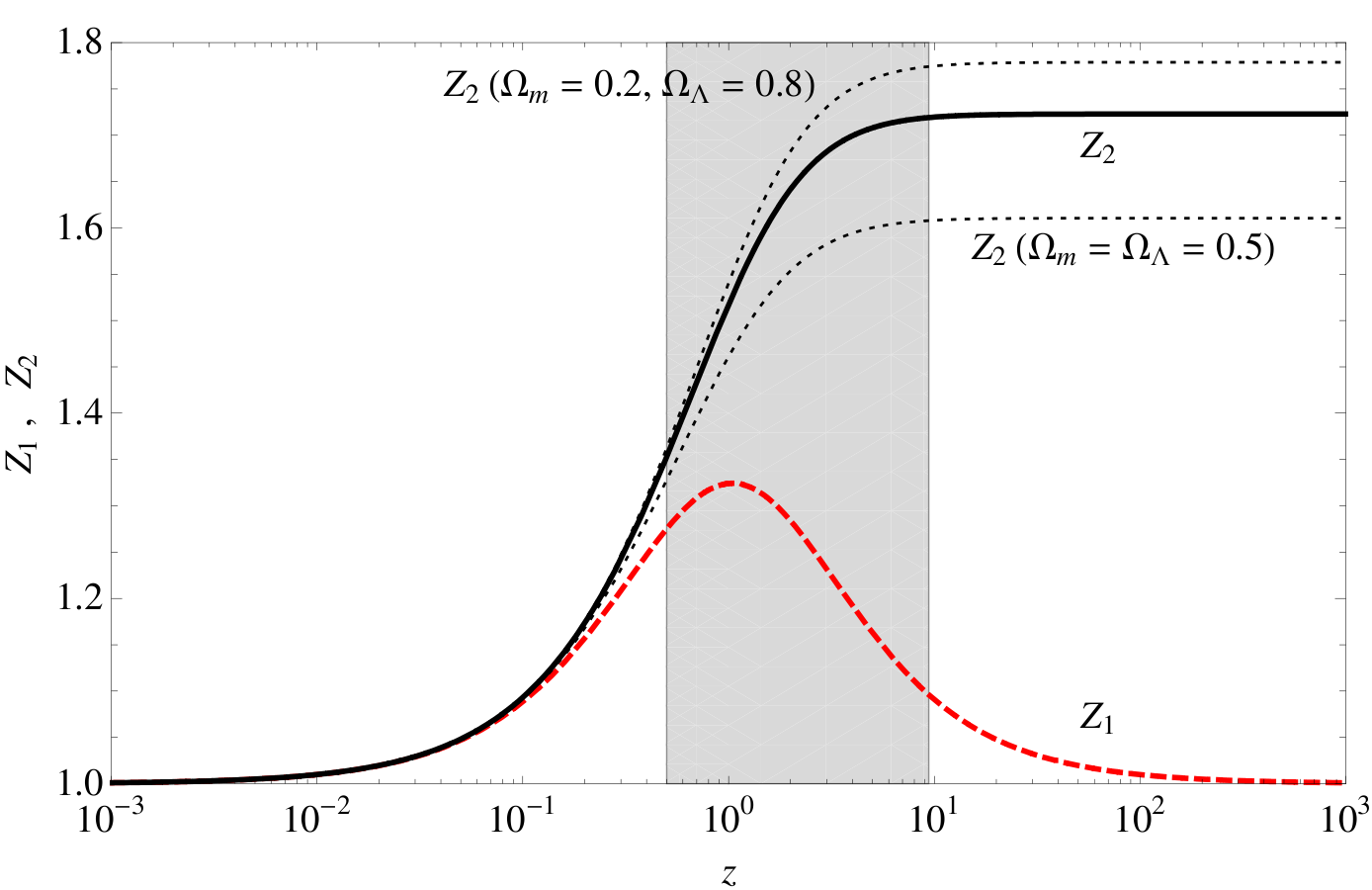}
 \mycaption{\label{fig:FigZDepComp}Comparison of the simplified ($\mathcal{Z}_1$) and proper ($\mathcal{Z}_2$) redshift dependencies. For $z \lesssim 0.1$ the approaches are identical. The gray-shaded region illustrates the typical redshift range populated by GRBs.}
\end{figure}
Since $\mathcal{Z}_1$ is a function of redshift, we show the functional dependence $\mathcal{Z}_1(z)$ in \figu{FigZDepComp}. Obviously, $\mathcal{Z}_1 \ge 1$, which means that the number of particles does not explode (third case excluded). However, for large $z$, $\mathcal{Z}_1 \rightarrow 1$, which means that the neutrinos from high redshifts are stable, and the notion of complete decays does not exist. From \equ{EqNz1}, they are even stable independently of the value of $E_0$, which is counter-intuitive, since more energetic neutrinos should live longer than less energetic ones. So what is wrong with this approach? Obviously, introducing the energy dependence \equ{EqEnergyRel} after the solution of \equ{EqDecay} does not take into account that the neutrinos lose energy continuously, which changes the solution of the differential equation. That is, the energy assigned to the neutrinos is chosen to be the one at the production point, which is on average too high, and the neutrinos are hence too long-lived.
While one would naively expect that this effect is small, we have encountered a severe contradiction here, which we are going to resolve in the next subsection.

\subsection{Solution of the redshift-dependent decay equation}
\label{sec:proper}

The proper solution is, of course, to re-write \equ{EqDecay} in terms of redshift using the redshift-dependent decay rate in \equ{EqDefLambda}:
\begin{equation}\label{equ:EqDefDNDz}
 \frac{dN_i(E_0,z)}{dz} = -\frac{\kappa_i}{E_0} \frac{dL}{dz} \frac{N_i(E_0,z)}{1+z}  \, .
\end{equation}
Note that in the following, we choose $z=0$ as the current epoch, \ie, we integrate from $z$ to $0$.
Therefore, from the definition of light-travel distance \equ{EqLDef}, we have
\begin{equation}\label{equ:EqDefDLDzlighttravel}
 \frac{dL}{dz} = - \frac{L_H}{\left(1+z\right)h\left(z\right)} \, .
\end{equation}
The solution is
\begin{equation}
 D_i(E_0,z) =\exp \left(-\frac{\kappa L_H}{E_0} \int_0^z \frac{dz^\prime}{\left(1+z^\prime\right)^2 h\left(z^\prime\right)} \right)
                           = \left[\mathcal{Z}_2\left(z\right)\right]^{-\frac{\kappa L_H}{E_0}} \, ,
\label{equ:damping}
\end{equation}
where we have isolated the $z$-dependent part through\footnote{Note that the function $I_2(z)$ was already encountered in the context of oscillations, and corresponds to Eq.~(6) in \Ref~\cite{Wagner:2011uy}.}
\begin{eqnarray}
 && I_2\left(z\right) \equiv \int_0^z \frac{dz^\prime}{\left(1+z^\prime\right)^2 h\left(z^\prime\right)} \label{equ:EqDefI2} \ , \\
 && \mathcal{Z}_2\left(z\right) \equiv e^{I_2\left(z\right)} \, . \label{equ:EqDefZ2}
\end{eqnarray}

The function $\mathcal{Z}_2\left(z\right)$ is also shown in \figu{FigZDepComp}. It is now monotonously increasing with redshift, and larger than unity. This means that for large decay rates $\kappa_i \gg E_0/L_H$, one finds $D_i(E_0,z) \rightarrow 0$ (complete decays) for $z>0$, as expected in a self-consistent framework. In addition, since the maximal distance is limited by the Hubble distance, the base $\mathcal{Z}_2$ in \equ{damping} is asymptotically limited. This has interesting implications for the notion of complete decays, which one can read off from \equ{damping}: The condition for complete decays is $\kappa L_H/E_0 \gg 1$, or, equivalently,
\begin{equation}\label{equ:EqCondCompDec}
 E_0 \, \left[\text{TeV}\right]  \ll \frac{4 \times 10^5}{\kappa^{-1} \, [\text{s eV}^{-1}]}  \, .
\end{equation}
First of all, this confirms our earlier estimates that energies considerably lower than 1~TeV are needed to test lifetimes comparable to the SN 1987A bounds. Second, it is a condition on energy, which is independent of distance (or redshift). Therefore, over cosmological distances (for $z \gtrsim 1$), ``complete decays'' is an energy-dependent concept, and has nothing to do with distance.

A useful analytical approximation for $\mathcal{Z}_2$ is
\begin{equation}\label{equ:EqDefZ2Approx}
 \mathcal{Z}_2\left(z\right) \simeq a + b e^{-cz} \, ,
\end{equation}
with\footnote{Actually, $a$ and $b$ are not independent, since $\mathcal{Z}_2(z=0)=1$ must be satisfied. This implies that $b = 1 - a$.} 
\begin{equation}\label{equ:EqDefABC}
 \left\{\begin{array}{l}
  a \simeq 1.71 \\
  b \simeq 1-a = -0.71 \\
  c \simeq 1.27
 \end{array} \right. ~.
\end{equation}
Hence, the asymptotic value is $\mathcal{Z}_2 \rightarrow a$ for $z \rightarrow \infty$. It depends on cosmology, as it is illustrated in \figu{FigZDepComp} for different values of $\Omega_m$ and $\Omega_\Lambda$. Clearly, the higher the cosmological constant contribution (for a flat universe), the more efficient neutrino decays will be for large redshifts. In principle, if the neutrino lifetime was known from different sources, one might even use this effect to independently probe cosmology by comparing low and high redshift populations of certain objects -- although the expected precision is probably not very high.

\section{Application to GRBs}
\label{sec:grb}

We introduce our benchmark GRBs and method in the first subsection. Then we show the impact of decay on the benchmark fluxes and compare the simplified and proper solutions. In the last subsection, we define a phenomenologically viable model and show the implications for fluxes and flavor ratios.

\subsection{Benchmark GRBs and method}

\begin{table}[t]
 \centering
 \begin{tabular}{|l|c|c|c|c|}
  \hline
                                         & SB                & GRB080916c          & GRB090902b          & GRB091024           \\
  \hline
  $\alpha$                               & 1                 & 0.91                & 0.61                & 1.01                \\
  $\beta$                                & 2                 & 2.08                & 3.80                & 2.17                \\
  $\epsilon_{\gamma,\text{break}}$ [MeV] & 1.556             & 0.167               & 0.613               & 0.081               \\
  $\Gamma$                               & $10^{2.5}$        & 1090                & 1000                & 195                 \\
  $t_v$ [s]                              & 0.0045            & 0.1                 & 0.053               & 0.032               \\
  $T_{90}$ [s]                           & 30                & 66                  & 22                  & 196                 \\
  $z$                                    & 2                 & 4.35                & 1.822               & 1.09                \\
  $\mathcal{F}_\gamma$ [erg cm$^{-2}$]   & $1 \cdot 10^{-5}$ & $1.6 \cdot 10^{-4}$ & $3.3 \cdot 10^{-4}$ & $5.1 \cdot 10^{-5}$ \\
  $L_\gamma^\text{iso}$ [erg s$^{-1}$]   & $10^{52}$         & $4.9 \cdot 10^{53}$ & $3.6 \cdot 10^{53}$ & $1.7 \cdot 10^{51}$ \\
  \hline
 \end{tabular}
 \mycaption{\label{tab:TblA} Properties of the four bursts discussed in the following, see \Ref~\cite{Baerwald:2010fk} for SB (``Standard Burst'', similar to \Refs~\cite{Waxman:1997ti,Waxman:1998yy}), \Refs~\cite{Nava:2010ig,Greiner:2009pm} for GRB080916c, \Refs~\cite{Nava:2010ig,Abdo:2009pg} for GRB090902b and \Refs~\cite{Nava:2010ig,Gruber:2011gu} for GRB091024. The luminosity is calculated with $L_\gamma^\text{iso}=4\pi d_L^2\cdot \mathcal{F}_\gamma/T_{90}$ with $\mathcal{F}_\gamma$ the fluence in the energy range $1\,\text{keV}-10\,\text{MeV}$. Adopted from \Ref~\cite{PhDHummer}.}
\end{table}

In the following, we apply the above decay framework to GRBs, some of which have been identified with high redshifts. Note, however, that our results can be equally applied to AGNs or other objects. We choose four benchmark bursts, for which the parameters are given in \Tab~\ref{tab:TblA}. In these cases, the observed photon spectrum is assumed to be a broken power law with spectral indices $\alpha$ and $\beta$ below and above the break at $\epsilon_{\gamma,\text{break}}$, respectively. The photon fluence is $\mathcal{F}_\gamma$ in the energy range $1\,\text{keV}-10\,\text{MeV}$. The estimated boost factor is $\Gamma$, the time variability is $t_v$, and the burst duration (where 90\% of the counts are observed) is $T_{90}$. As far as the selection of bursts is concerned, the standard burst ``SB'' is a burst for which most parameters, such as $\alpha$, $\beta$, $z$, and $L_\gamma^\text{iso}$, are frequently used in the literature. In addition, three different recent Fermi-measured GRBs have been chosen as examples: GRB 080916C, GRB 090902B, and GRB 091024. GRB 080916C has been selected, because it is one of the brightest bursts ever seen, although at a large redshift, and one of the best studied Fermi-LAT bursts. The gamma-ray spectrum of GRB 090902B has a relatively steep cutoff. GRB 091024 can be regarded as a typical example representative for many Fermi-GBM bursts~\cite{Nava:2010ig}, except for the long duration. Note that the middle two bursts have an exceptionally large $\Gamma \gtrsim 1000$, whereas $\Gamma \simeq 200$ for the last burst. All three observed bursts have in common that the required parameters for the neutrino flux computation can be taken from the literature; see the figure caption for the references. 

Once the parameters in \Tab~\ref{tab:TblA} are given, the neutrino fluxes can be predicted, using methods such as in \Refs~\cite{Guetta:2003wi,Abbasi:2009ig,Hummer:2011ms}. We use the most updated numerical method from \Ref~\cite{Hummer:2011ms} (see \Ref~\cite{Baerwald:2011ee} for details), which has corrected earlier flux predictions down by about one order of magnitude by taking into account the full spectral information, the energy dependence of the mean free path of the protons, and the cooling of secondaries including adiabatic energy losses. In addition, $t$-channel and high-energy photo-pion production processes, neutrinos from muon, pion, kaon, and neutron decays, the helicity dependence of the muon decays, and flavor mixing are included. As one of the additional assumptions of the method, we assume that the protons carry ten times more energy than the electrons, and that there is energy equipartition between electrons and magnetic field energy, \ie, we use the same parameters as in \Refs~\cite{Abbasi:2009ig,Abbasi:2011qc,Abbasi:2012zw}. 
With these parameters it is possible to estimate the maximal proton energy by balancing the acceleration rate of protons (acceleration efficiency $0.1$ used) with the faster of the synchrotron loss and adiabatic loss rates, which are assumed to be the dominant energy loss rates. 

\subsection{Impact of decay on neutrino fluxes, and comparison of methods}

\begin{figure}[tp]
  \centering
  \includegraphics[width=0.8\textwidth]{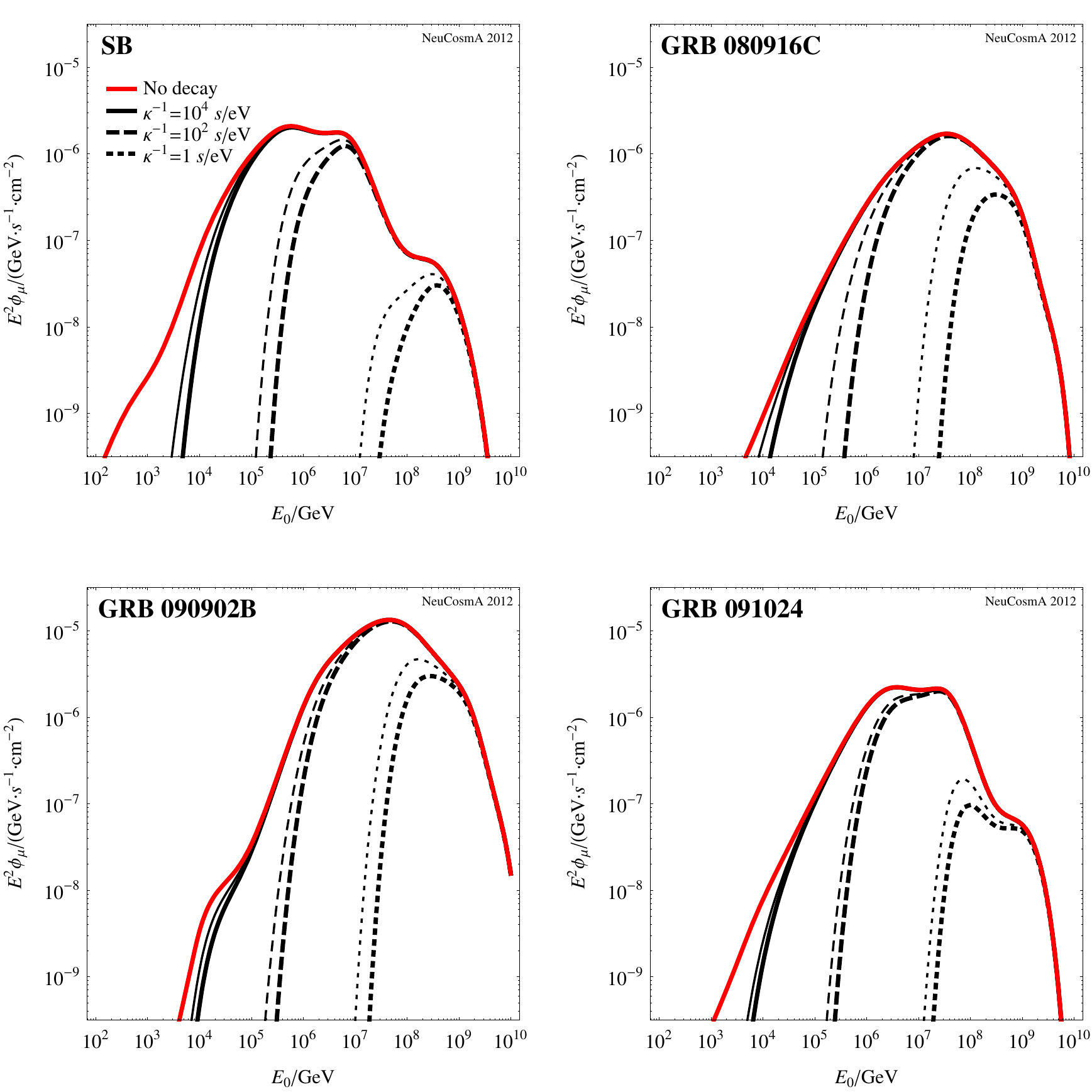}
\mycaption{\label{fig:complete}
Predicted muon neutrino flux as a function of neutrino energy for the four benchmark bursts in \Tab~\ref{tab:TblA} (panels). The different curves in each panel correspond to no decay (stable neutrinos) and three different lifetimes, as given in the legend. Thick curves correspond to the proper solution in \Sec~\ref{sec:proper}, and thin curves to the simplified approach in \Sec~\ref{sec:simple} for comparison. Here we (hypothetically) assume that all mass eigenstates have the same lifetime, as indicated in the legend.
}
\end{figure}

In order to show the impact of neutrino decay on the neutrino fluxes, let us first of all choose a (hypothetical) model where all mass eigenstates have the same lifetimes. We show in \figu{complete} the predicted muon neutrino flux\footnote{Here the neutrino flux is just the neutrino fluence divided by $T_{90}$.} as a function of neutrino energy for the four benchmark bursts in \Tab~\ref{tab:TblA} for this model. The different lifetimes used for the different curves are shown in the plot legend. Thick curves correspond to the proper solution in \Sec~\ref{sec:proper}, and thin curves to the simplified approach in \Sec~\ref{sec:simple} for comparison. Note that this model is hypothetical in the sense that the lifetimes are chosen such that the SN 1987A bound for one of the mass eigenstates must be violated. However, it is the simplest model to illustrate the implications of neutrino decay.

The actual shapes of the different predictions in the stable case (``no decay'') depend on the burst parameters. For large enough magnetic fields, the characteristic wiggles from the secondary (pion, muon, kaon) cooling can be seen. For instance, for ``SB'' (upper left panel), the three peaks correspond, from left to right, to muon decays, pion decays, and kaon decays, respectively. In some cases, the peaks are less pronounced, especially if adiabatic cooling on protons and secondaries is important (such as for GRB 080916C); \cf, Fig.~13 in \Ref~\cite{Winter:2012xq}.

As far as the impact of neutrino lifetime is concerned, note that all bursts in \Tab~\ref{tab:TblA} have redshifts larger than one. As a consequence, we can use \equ{EqCondCompDec} to estimate where the decays will be complete. In fact, the energy $E_0$, shown in the plot, below which the flux is practically switched off, can be computed with this equation. For instance, in the upper left panel, the flux is suppressed below $4 \, 000 \, \mathrm{TeV}$ for $\kappa^{-1} = 10^2 \, \mathrm{s} \, \mathrm{eV}^{-1}$. This condition is the more accurate, the higher the redshift of the burst is. For instance, for GRB 091024 ($z \simeq 1.1$), the suppression is somewhat weaker (shifted to smaller energies) because the base $\mathcal{Z}_2$ in \figu{FigZDepComp} has not yet reached its saturation value. 

Let us now compare the simplified approach in \Sec~\ref{sec:simple} (thin curves) with the proper solution in \Sec~\ref{sec:proper} (thick curves): Obviously the suppression in the proper treatment is stronger, which is expected from $\mathcal{Z}_2 \gg \mathcal{Z}_1$ for large redshifts. The difference between the curves also increases with redshift, and can be quite significant for large redshifts (\cf, GRB 080916C). Now one may argue that for the individual burst, one would measure a too short lifetime in the simplified approach, which just needs to be corrected. This argument, however, breaks down if different bursts at different redshifts are compared. Let us illustrate this with one example.
For the diffuse neutrino flux, it is expected that the peak contribution comes from a redshift $z \simeq 1$~\cite{Baerwald:2011ee} (Fig.~3), which means that low redshift ($z < 1$) and high redshift ($z > 1$) bursts contribute almost equally. In a diffuse (or a stacked) neutrino flux, the fluxes from these individual bursts are summed over.  For $z \ll 1$, the simplified and proper approaches are identical, whereas for $z \gtrsim 1$, the properly calculated fluxes are more strongly suppressed for a certain lifetime. Hence, if the simplified approach is used, there will be a contradiction implied.

It is, in fact, also interesting to study if  $z \simeq 1$ as the dominant contribution actually holds in the presence of decay. We have tested this for different values of $E_0$. For instance, we choose $E_0 \simeq 10^6 \, \mathrm{GeV}$, which is the peak for the SB spectrum, where the neutrino telescopes typically have good sensitivity. \equ{EqCondCompDec} can then be used to estimate that the flux at this energy is suppressed for $\kappa^{-1} \lesssim 400 \, \mathrm{s} \, \mathrm{eV}^{-1}$. From the convolution of the GRB rate (following the star formation rate, possibly with some correction), the $1/d_L^2$ drop of $E^2 \phi$, and the decay damping, one can show that at this energy, the main contribution does not come from $z \simeq 1$ anymore, but from very low $z$ (peaking at $z=0$) if $\kappa^{-1} \lesssim 400 \, \mathrm{s} \, \mathrm{eV}^{-1}$.  This is expected, since the neutrinos from the low redshift bursts will not have decayed yet. If, on the other hand, $\kappa^{-1} \gtrsim 400 \, \mathrm{s} \, \mathrm{eV}^{-1}$, the bursts from $z \simeq 1$  will dominate, as for stable neutrinos.

Finally, comparing the lifetimes used for \figu{complete} with the SN 1987A bound, one can easily see that no strong decay suppression can be obtained if this bound is applied to all mass eigenstates. Therefore, the neutrino lifetimes of the different mass eigenstates have to be different for a phenomenologically allowed scenario, where the most plausible case, satisfying the SN 1987A constraint, might be that $\nu_1$ is stable. In fact, this is for the discussed GRB fluxes equivalent to saturating the SN 1987A bound on $\nu_1$. Therefore, we only discuss this case in the following, and we only use the proper solution of the decay equation.

\subsection{Fluxes and flavor ratios for $\boldsymbol{\nu_1}$ stable}
\label{sec:m1stable}

As discussed above, we focus in the following only on the case of $\nu_1$ stable, which satisfies the SN 1987A constraint. Recall that the reason for this constraint is that only the electron flavor has been measured by the SN 1987A observation, which consists mainly of $\nu_1$. On the other hand, we assume that $\nu_2$ and $\nu_3$ decay with the same lifetime, for the sake of simplicity, with lifetimes satisfying the $\nu_2$ (and, consequently, $\nu_3$) bound. The mass eigenstate $\nu_2$ is about an equal mixture among $\nu_e$, $\nu_\mu$, and $\nu_\tau$, whereas $\nu_3$ is a mixture between $\nu_\mu$ and $\nu_\tau$ with a small contribution of $\nu_e$. The decay of $\nu_2$ will affect the electron flavor somewhat, which however can be absorbed in the uncertainties of current supernova flux predictions. The decay of $\nu_3$ will be hardly observable in the electron flavor.

\begin{figure}[tp]
  \centering
  \includegraphics[width=0.8\textwidth]{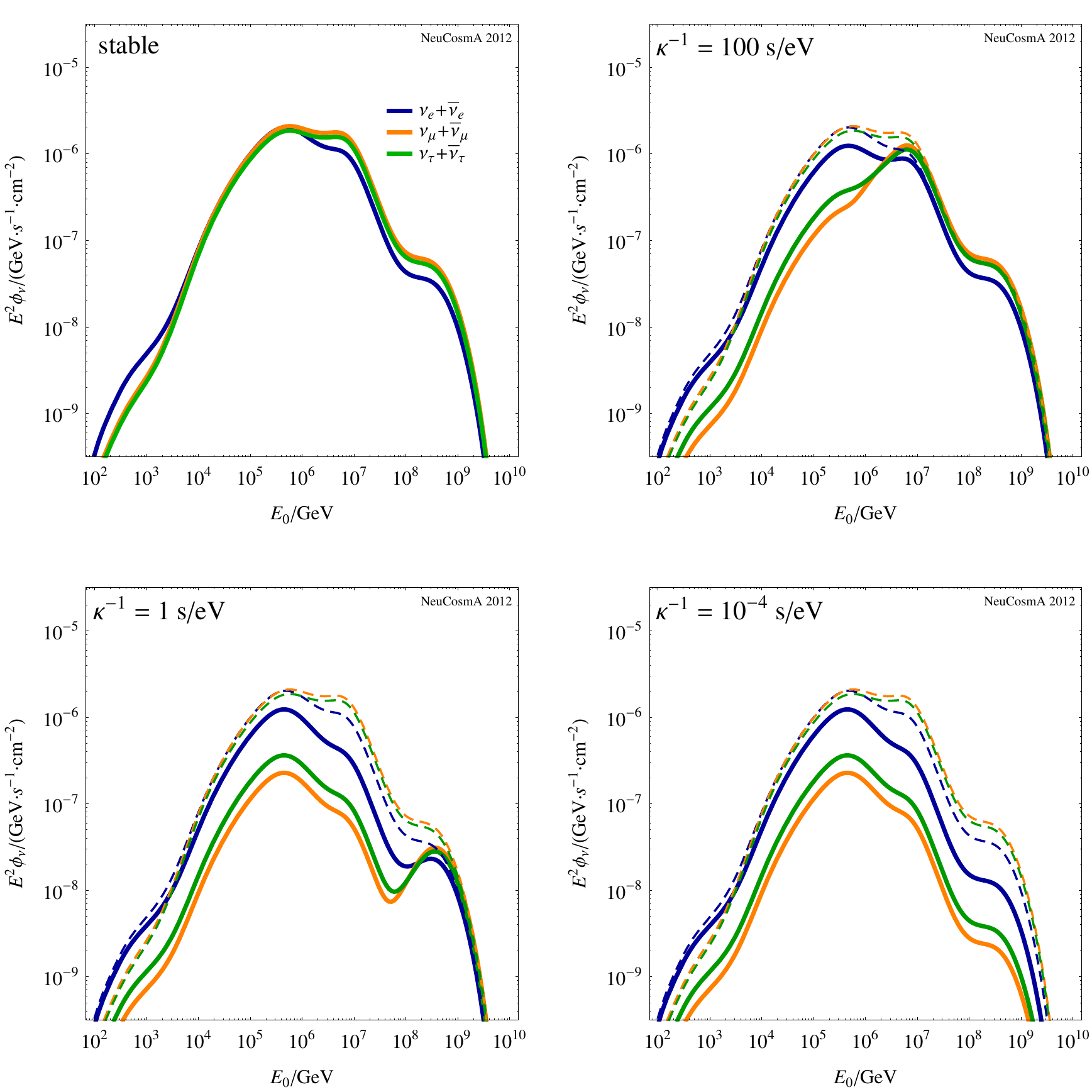}
\mycaption{\label{fig:lifetime}
Predicted muon neutrino flux as a function of neutrino energy for the benchmark burst SB in \Tab~\ref{tab:TblA}.
The different panels correspond to different lifetimes, as given in the panels. The different curves correspond to different flavors (after flavor mixing), as given in the legend. Thin dashed curves shown the stable case (upper left panel) for reference.  As decay model, $\nu_1$ is assumed to be stable, whereas $\nu_2$ and $\nu_3$ decay with the indicated lifetime, \ie, all of the panels are consistent with the SN 1987A bound.
}
\end{figure}

We show the electron, muon, and tau neutrino fluxes for this model in \figu{lifetime} for the standard burst SB and different values of the lifetime (panels). In the stable case (upper left panel) at the peak, where the flavor composition $\nu_e:\nu_\mu:\nu_\tau$ is approximately $1:2:0$ at the source, flavor equilibration at the detector is roughly achieved. Even for larger energies, the fluxes of all flavors are roughly comparable on a logarithmic scale -- we will discuss the flavor ratio below. In the most extreme decay case (lower right panel), both $\nu_2$ and $\nu_3$ decay with a lifetime corresponding to the current $\nu_2$ bound. In this case, the electron neutrino flux survives (because $\nu_1$ is stable), whereas the muon and tau neutrino fluxes are suppressed. The asymmetry between the muon and tau neutrino fluxes mainly comes from the large value of $\theta_{13}$, the chosen $\deltacp$, and the non-maximal $\theta_{23}$. 
As a consequence:
\begin{itemize}
\item
 Muon tracks are suppressed by about one order of magnitude compared to the stable case.
\item
 Electromagnetic cascades are observable at about the same level as in the stable case.
\item
 Hadronic cascades are also strongly suppressed. The contributions of the leptonic tau decay channel into leptons can be neglected as well, since they are doubly suppressed by tau neutrino flux and the tau branching ratios.
\item
 Neutral current cascades are reduced to about $1/3$ of the stable case.
\item
 Glashow resonances should occur at roughly the expected level as in the stable case. Note, however, that the right energy has to be hit (6.3~PeV), and that Glashow resonant events are expected to be suppressed for neutrino production from photohadronic interactions, see, \eg, \Ref~\cite{Hummer:2010ai} for a detailed (quantitative) discussion.
\end{itemize}
It is therefore clear that this scenario can easily explain why only cascades have been recently observed at PeV energies, and will be dominating with increasing statistics, in spite of the effective areas comparable between muon tracks and cascades~\cite{NeutrinoTalk}. The PeV energies are exactly at the peak expected from the standard burst (SB), which means that the events may come from the diffuse GRB flux.

From the lower left panel of \figu{lifetime}, one can read off that the qualitative conclusions do not change even if the $\nu_2$ and $\nu_3$ lifetimes are increased by four orders of magnitude. In this case, only the rightmost peak (neutrino production from kaon decays) is not affected by the decays. A very interesting case are intermediate lifetimes, such as $\kappa^{-1} \sim 100 \, \mathrm{s} \, \mathrm{eV}^{-1}$ in the upper right panel. Here the muon (and tau) neutrino flux dominates above about 1~PeV, whereas below 1~PeV the electron neutrino flux dominates, \ie, there is a spectral swap induced by neutrino decay.

\begin{figure}[tp]
  \centering
  \includegraphics[width=1.0\textwidth]{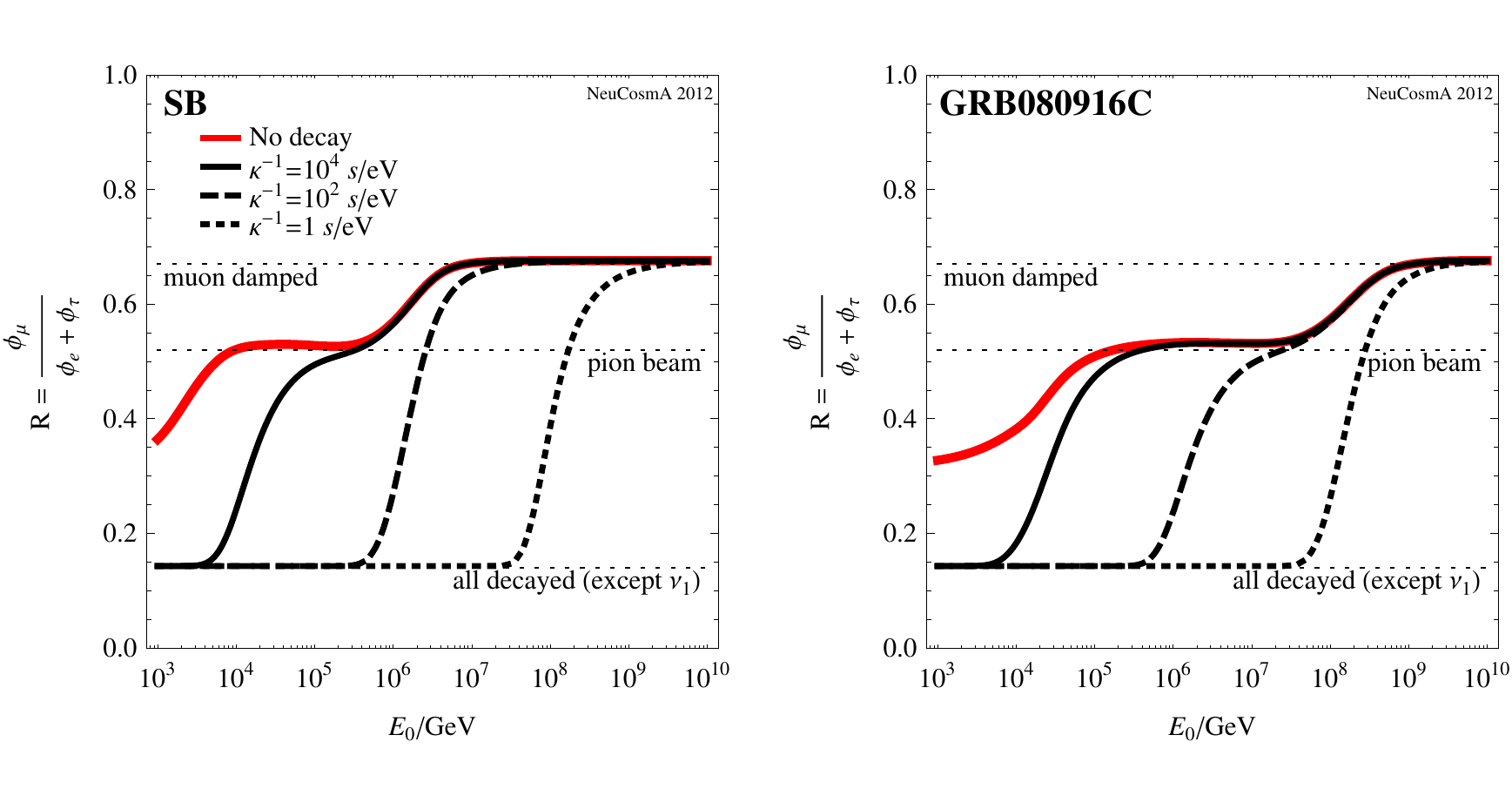}
\mycaption{\label{fig:flratio}
Predicted flavor ratio $R$ (muon tracks to cascades) as a function of neutrino energy for the first two  benchmark bursts in \Tab~\ref{tab:TblA} (panels). The different curves in each panel correspond to no decay (stable neutrinos) and three different lifetimes, as given in the legend.  As decay model, $\nu_1$ is assumed to be stable, whereas $\nu_2$ and $\nu_3$ decay with the indicated lifetime, \ie, all of the lifetimes are consistent with the SN 1987A bound.
}
\end{figure}

The implications for the flavor composition can be better seen on a linear scale. We therefore show in \figu{flratio} the  flavor ratio $R$ (muon tracks to cascades) as a function of neutrino energy for two different bursts (panels) and different lifetimes (curves). One can easily see the characteristic transition from a pion beam source (initial flavor composition $\nu_e$:$\nu_\mu$:$\nu_\tau$ of $1:2:0$, yielding $R \approx 0.52$) to a muon damped source (initial flavor composition of $0:1:0$, yielding $R \approx 0.67$) at higher energies, which comes from the energy losses of the muons in the magnetic field. It depends, apart from particle physics parameters, on the value of $B'$, the magnitude of the magnetic field in the shock rest frame; see \Ref~\cite{Winter:2012xq} for a review. The different panels show two different cases for this transition.

It is easy to show that if only one mass eigenstate is stable, an asymptotic value of the ratio $R$ is reached for complete decays which does not depend on the initial flavor composition (see, \eg, \Refs~\cite{Pakvasa:1981ci,Maltoni:2008jr}). This asymptotic value $R \simeq 0.14$ is marked in \figu{flratio}, and it is always reached for low enough energies in the presence of decay. The low value simply reflects that muon tracks are strongly suppressed compared to the cascades.  The transition towards this asymptotic curve depends on the lifetime, of course. Depending on the chosen lifetime and magnetic field in the source, the pion beam flavor ratio may be actually observable in a certain energy range, such as for $\kappa^{-1} = 10^4 \, \mathrm{s} \, \mathrm{eV}^{-1}$ in the left panel, or not, such as for $\kappa^{-1} = 10^2 \, \mathrm{s} \, \mathrm{eV}^{-1}$ in the left panel. The presence of different flavor ratios may be useful to distinguish different decay scenarios; see \Refs~\cite{Maltoni:2008jr,Mehta:2011qb}.

\section{Consequences for realistic GRB stacking analyses}
\label{sec:icecube}

\begin{figure}[t]
  \centering
  \includegraphics[width=\textwidth]{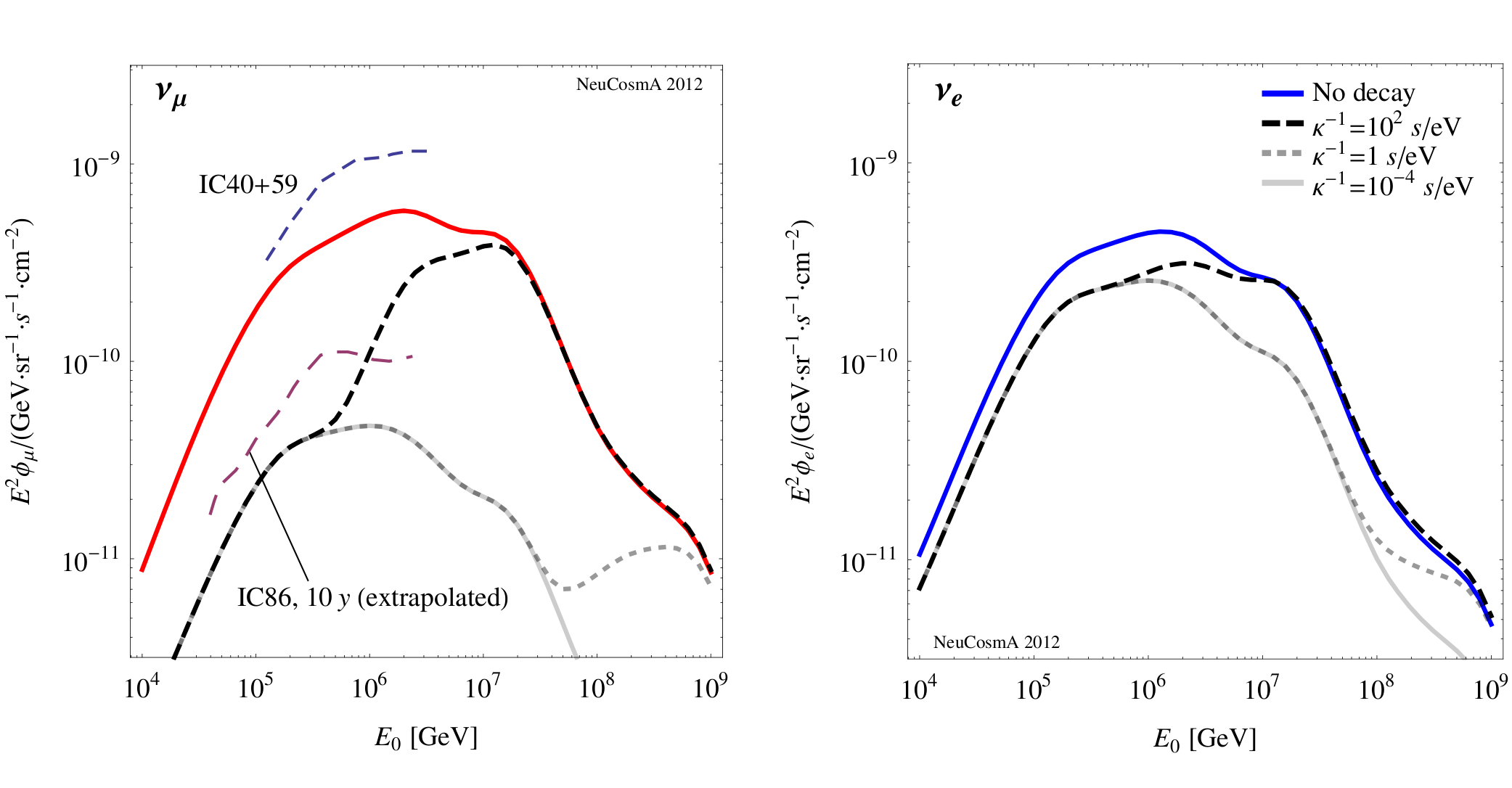}
\mycaption{\label{fig:icecube} 
Prediction of the quasi-diffuse muon (left panel) and electron (right panel) neutrino flux from the bursts used in the IceCube stacking analysis. The different curves in each panel correspond to no decay (stable neutrinos) and three different lifetimes, as given in the legend, which are applied to each burst in the stacking sample individually.  As decay model, $\nu_1$ is assumed to be stable, whereas $\nu_2$ and $\nu_3$ decay with the indicated lifetime, \ie, all of the lifetimes are consistent with their current bounds. The limit from IC-40+59 is shown~\cite{Abbasi:2012zw}, as well as an extrapolated one for IC-86 (ten years, $A_{\mathrm{eff}}^{\mathrm{IC86}} \simeq 3 \times A_{\mathrm{eff}}^{\mathrm{IC40}}$; see, \eg, \Ref~\cite{Karle:2010xx}). For the prediction, the 117~bursts in the IC-40 sample have been used~\cite{Abbasi:2011qc}, the ``no decay'' corresponds to the nominal prediction in Fig.~3 of \Ref~\cite{Hummer:2011ms}. 
}
\end{figure}

We now apply the scenario in \Sec~\ref{sec:m1stable} to a realistic sample of GRBs, as it has been actually used in a stacking analysis. Remember that from the SN 1987A constraint, we have been forced to choose $\nu_1$ stable for all practical purposes, whereas we have assumed that $\nu_2$ and $\nu_3$ decay with equal rates. 

We show in \figu{icecube} the prediction for the quasi-diffuse muon (left panel) and electron (right panel) neutrino fluxes from the 117~bursts in the IC-40 stacking sample as curves labeled ``no decay''. The left curve  corresponds to the nominal prediction in Fig.~3 of \Ref~\cite{Hummer:2011ms}, updated with the current values of the mixing angles. It uses exactly the same bursts and parameters as the original IceCube analysis, but the numerical (updated) method for the flux prediction from \Ref~\cite{Hummer:2011ms}.  Note that the quasi-diffuse flux normalization hardly depends on the burst sample used, but only the shape (\cf, dashed curves in Fig.~1 of \Ref~\cite{Abbasi:2012zw}). This means that the shape will be affected by the actual bursts in the sample, whereas the normalization will remain approximately constant -- modulo statistical fluctuations from the quasi-diffuse flux extrapolation; see \Refs~\cite{Halzen:1999xc,Baerwald:2011ee}. We also show in \figu{icecube}, left panel, the current stacking limit from muon tracks, as well as an extrapolation for the full-scale experiment. While the current limit does not yet exceed our nominal prediction, the full scale experiment will finally exceed it for the chosen standard values of the astrophysical parameters. Note that in the original IceCube analysis, $z=2$ has been assigned to a burst if the redshift has not been measured, while the pion production efficiency has been computed (for long bursts) with $L_\gamma^{\mathrm{iso}}=10^{52} \, \mathrm{erg} \, \mathrm{s}^{-1}$. Since we stick to these rules, most bursts will have a redshift $z > 1$, and cosmological effects will be important for decay.

In \figu{icecube}, we also show the flux predictions for different neutrino lifetimes. In the extreme cases, $\kappa^{-1} \lesssim 1 \,\mathrm{s} \, \mathrm{eV}^{-1}$, the muon neutrino predicted flux is suppressed to below the expected limit from IC-86 over ten years. However, in these cases, a substantial electron neutrino flux is expected; see right panel. It can be read off from the figure, that gamma-ray burst neutrinos should be detected with the full scale experiment in cascades if the effective area for cascades is at least 25\% of the one  for muon tracks at 1~PeV.\footnote{Unfortunately, there are no GRB cascade analyses available up to this point, which means that reliable limits in the right panel of \figu{icecube} cannot be shown.} On the other hand, no reliable information on astrophysical neutrino sources, such as GRBs, can be obtained from muon tracks only. In particular, it is not clear if no neutrinos are found because they decay, or because the baryonic loading in GRBs is smaller than anticipated. 

A very interesting case is the intermediate value $\kappa^{-1} \simeq 100 \,\mathrm{s} \, \mathrm{eV}^{-1}$ and its implications for muon tracks; see left panel. In this case, only the low energy component of the flux is damped, and only neutrinos above a certain threshold energy are observable. It is expected that the full scale experiment may detect neutrinos in such  a case, but the interplay with the (energy-dependent) detector response becomes important (see \Ref~\cite{Winter:2011jr} for a more detailed discussion). For instance, better information may be obtained from down-going than up-going muon tracks, because these exhibit a better sensitivity at higher energies (\cf, Fig.~3 in  \Ref~\cite{Winter:2011jr}). As a consequence, the sensitivity may be dominated by a different viewing window and therefore different bursts compared to the ``no decay'' case, which may have different characteristics. One has to keep that in mind if one compares the sensitivities for these cases.

\section{Summary and discussion}
\label{sec:summary}

We have discussed neutrino decays of ultra-relativistic neutrinos over cosmological distances, \ie, for redshifts $z \gtrsim 1$. These neutrinos may originate from extragalactic cosmic accelerators, such as AGNs or GRBs. As class of  decay models, we have considered decays into secondaries invisible to the detector, such as Majorons, unparticles, sterile neutrinos, or neutrinos strongly degraded in energy. The decays have been allowed to be incomplete, \ie, the spectral signature of the decay may be seen. 

We have demonstrated that simplified approaches to neutrino decay cannot be used over such distances, and that the differential decay equation has to be carefully solved as a function of redshift. We have also given a simple parameterization for the solution, which can be easily used. From a comparison of different approaches, we have demonstrated that conventional approaches to neutrino decay can be safely used for $z \lesssim 0.1$ (or distances $L \lesssim 360 \, \mathrm{Mpc}$), whereas the light-travel (or look-back) distance triggers the clock for neutrino decays over cosmological distances. As a consequence, the horizon is limited by the Hubble length $\simeq 3.89 \, \mathrm{Gpc}$, which is the maximal distance scale testable by decay. This has several interesting consequences: First of all, the common notion ``the longer the distance, the longer lifetimes can be tested'' does not hold anymore. Second, the concept of complete decays (all neutrinos have decayed) cannot be assigned to long enough distances, but only to low enough energies, \ie, it is an energy-dependent and not distance-dependent statement. And third, the neutrino lifetime bound from SN 1987A, which emitted neutrinos at much lower energies, cannot be exceeded by high-energy astrophysical neutrinos from AGNs and GRBs tested in neutrino telescopes. The same argument applies to the cosmogenic neutrinos from cosmic microwave (and infrared) background interactions, which are at even higher energies. Note that the SN 1987A neutrinos have only been measured in the  electron flavor, which means that this limit does not apply to all mass eigenstates. However, it is expected that the neutrinos from any future galactic supernova explosion will be measured in different flavors and neutral currents, which means that in long terms the best lifetime bounds are expected to come from supernova neutrinos.

While our results can be easily applied to AGNs as well, we have used GRBs as test case. Because of the SN 1987A constraint, we have been forced to choose $\nu_1$ stable and $\nu_2$ and $\nu_3$ decaying with equal rates.\footnote{We have demonstrated that saturating the SN 1987A bound for the $\nu_1$ lifetime does not change the results, which means that it is equivalent to the studied model for the purpose considered.}
We have shown that, safely within the current bounds for $\nu_2$ and $\nu_3$, the muon neutrino flux can be substantially reduced  by about one order of magnitude, whereas the electron neutrino flux is hardly affected. This has very interesting implications for neutrino telescopes: First, improved bounds on the neutrino flux from muon tracks can be interpreted in terms of the astrophysical parameters or the possibility that neutrinos are unstable. On the other hand, the electromagnetic cascade bounds are hardly affected by the neutrino lifetime, which means that reliable conclusions on the astrophysical objects can only come from cascades, and highlights the importance of dedicated cascade analyses for GRBs and other classes of objects. This would be different, of course, if muon tracks from astrophysical neutrinos were actually observed, since their discovery would constrain this possibility. 

For GRBs, in particular, we have used a realistic sample of IceCube GRBs and the state-of-the-art technology  for the prediction of the quasi-diffuse neutrino flux to illustrate the consequences. We have demonstrated that, for the discussed decay scenarios and our nominal flux prediction using the same parameters as IceCube, even the full scale IceCube experiment may not find neutrinos from GRBs in muon tracks after ten years, whereas the perspectives for a detection in cascades are actually very good. 

Finally, note that recently two cascade neutrino event candidates at PeV energies have been found.  Our scenario provides a plausible explanation why only cascades are seen, whatever the origin of the neutrinos is, whereas there are no accompanying muon tracks in spite of comparable effective areas. For GRBs, in particular, the PeV energies are exactly where we predict the maximum of the quasi-diffuse flux; see \figu{icecube}, right panel. On the other hand, for the relevant search time window, we only expect 0.07 electromagnetic cascade events for the nominal (``no decay'') prediction from GRBs. The observed two events might in this case come from a strong statistical fluctuation, or a significant deviation of astrophysical parameters from their assumed mean values.

\acknowledgments

We would like to thank Markus Ahlers, Amol Dighe, Francis Halzen, Svenja H{\"u}mmer, Poonam Mehta, Irina Mocioiu, Sandip Pakvasa, Eli Waxman, and Nathan Whitehorn for useful discussions on aspects regarding this work. 

We would like to acknowledge support from DFG grants WI 2639/3-1 and
WI 2639/4-1.  MB and PB would like to acknowledge support from the GRK 1147 ``Theoretical Astrophysics and Particle Physics''. WW would
like to thank GGI Florence for hospitality during their stay within
the ``What's $\nu$?'' program. 
This work has also been supported by the FP7 Invisibles network (Marie Curie
Actions, PITN-GA-2011-289442) and the ``Helmholtz Alliance for Astroparticle Physics HAP'', funded by the Initiative and Networking fund of the Helmholtz association.

% \bibliographystyle{h-physrev5}
% \bibliography{references}

\end{document}